\documentclass[showpacs,amsmath,amssymb,aps,pre]{revtex4}

\usepackage{graphicx}
\usepackage{dcolumn}
\usepackage{bm}

\begin{document}

\preprint{}

\title{Transport in a highly asymmetric binary fluid mixture}

\author{Sorin Bastea}
\email{sbastea@llnl.gov}
\affiliation{Lawrence Livermore National Laboratory, P.O. BOX 808, Livermore, CA 94550}


\begin{abstract}
We present molecular dynamics calculations of the thermal conductivity and 
viscosities of a model colloidal suspension with colloidal particles roughly 
one order of magnitude larger than the suspending liquid molecules. The results are 
compared with estimates based on the Enskog transport theory and effective 
medium theories (EMT) for thermal and viscous transport. We find, in particular, 
that EMT remains well applicable for predicting both the shear viscosity and 
thermal conductivity of such suspensions when the colloidal particles have a 
``typical'' mass, i.e. much larger than the liquid molecules. Very light 
colloidal particles on the other hand yield higher thermal conductivities, 
in disagreement with EMT. We also discuss the consequences of 
these results to some proposed mechanisms for thermal conduction in 
nanocolloidal suspensions. 
\end{abstract}

\pacs{66.60.+a, 66.20.+d, 82.70.Dd, 83.80.Hj}

\maketitle

Liquid suspensions of solid particles (colloids) are widely encountered in biology, 
industry and many natural processes. In addition to their relevance 
for numerous practical applications they have emerged as a useful paradigm 
for the study of phase transitions, from crystal nucleation and growth to 
gelation \cite{al02}. Colloids have not only interesting thermodynamic 
properties, but remarkable rheological properties as well \cite{ssc93} and 
very complex flow behavior \cite{mdh04}. When the suspended particles are only 
one to two orders of magnitude larger than typical liquid molecules, i.e. in the 
nanometer domain, colloids may exhibit entirely new properties \cite{wn03,kpc02} 
that are expected to have important technological consequences. 
A theoretically important and practically relevant class of colloids 
consists of suspensions of spherical colloidal particles with interactions 
dominated by excluded volume effects. In the following we employ molecular 
dynamics simulations 
to calculate the thermal conductivity and viscosities (shear and bulk) of fairly 
dilute colloidal suspensions modeled as mixtures of strongly asymmetric spherical 
particles 
interacting through short range repulsive potentials. We discuss the results in 
the light of theoretical estimates based on microscopic and macroscopic 
pictures of the system. Despite the simplicity of the model the conclusions 
should provide, {\it inter alia}, 
some guidance on the expected transport properties of dilute suspensions of 
nanosized particles, particularly the thermal conductivity, which has been 
the subject of some speculation \cite{kpc02,dhk04,sb05}.  

The model that we study consists of two types of particles, {\it 1} - {\it solvent} 
and {\it 2} - {\it colloid}, with masses $m_1\equiv m$ and $m_2\equiv m_c$. The interaction potentials 
between the particles are based on the inverse-12, ``soft sphere'' potential, whose 
properties have been well studied  \cite{hrj70,ah75}
\begin{equation}
u(r)=\epsilon\left(\frac{\sigma}{r}\right)^{12}
\end{equation}
and which we truncate and shift at $r/\sigma=2$; we also define $u(r)=\infty$ for $r<0$. 
The interactions are:
\begin{eqnarray}
&&u_{11}(r) = u(r)\\
&&u_{12}(r) = u(r - R_c)\\
&&u_{22}(r) = u(r - 2R_c)
\end{eqnarray}
Similar interactions, that take into account 
the "size" of the colloidal particles by 
introducing a "hard core radius" $R_c$, have been employed before to model suspensions 
\cite{nmh98,sb06}. For temperatures $k_B T\simeq \epsilon$ the effective diameters corresponding to 
the above interactions should be well approximated by $\sigma_1=\sigma$, 
$\sigma_{12}=R_c+\sigma$, and $\sigma_2\equiv\sigma_c=2R_c + \sigma$, and 
satisfy $\sigma_{12} = (\sigma_1 + \sigma_2)/2$. In the following we will therefore quote 
as relevant quantities the "diameter" ratio of the colloid and solvent particles, 
$\sigma_c/\sigma$, and the "volume fractions" of the colloidal particles, 
$\phi_c=\pi n_c{\sigma_c}^3/6$, and the solvent, $\phi=\pi n{\sigma^3}/6$; $n_c$ and $n$ 
are the corresponding number densities, $n_c=N_c/V$, $n=N/V$. All the simulations 
presented here were performed in the microcanonical (NVE) ensemble with the average 
temperature set to $k_B T=\epsilon$.

The systems (mixtures) that we studied are summarized in Table 1, and correspond to two 
"diameter" ratios, each with two mass ratios. Since for a realistic colloidal particle the 
ratio of its mass to that of a fluid molecule is $m_c/m\sim (\sigma_c/\sigma)^3$, our perhaps 
most practically relevant results correspond to $m_c/m=1000$. However, we also analyzed 
the effect of a much 
smaller mass ratio, $m_c/m=1$. The volume fractions $\phi_c$ of the colloidal particles have 
been chosen low enough so that the system is 
rather dilute, but high enough so that a reasonable number of colloidal particles 
can be simulated without the need for a prohibitively large number of solvent particles. 
(Nevertheless, $N$ is rather large, of the order $10^5$.) The pairs $\phi$ and $\phi_c$ for the 
two different diameter ratios have been chosen to yield the same system pressure $p_0$, corresponding 
to a pure solvent
at $n_0\sigma^3=0.8$ $(\phi_0=0.419)$. Incidentally, we have found that this can be accomplished with very 
good precision (better than $1\%$) by using the scaling relation:
\begin{equation}
\frac{p(\sigma, \sigma_c, \phi, \phi_c)}{p_0(\sigma, \phi_0)}=
\frac{p_{MCSL}(\sigma, \sigma_c, \phi, \phi_c)}{p_{CS}(\sigma, \phi_0)}
\end{equation}
where $p$ is the system pressure, $p_{BMCSL}$ the Boubl\'{i}k-Mansoori-Carnahan-Starling-Leland 
equation of state pressure of a hard-sphere mixture \cite{tb70,mcsl} and $p_{CS}$ the Carnahan-Starling 
equation of state pressure of the hard-sphere liquid \cite{hm}. The choice of a common 
pressure allows an unambiguous comparison of the 
thermal conductivity and viscosities of the suspension with that of the reference system (pure 
solvent at pressure $p_0$), is implicitly assumed by theories relying on macroscopic scale 
arguments (see below), and should also correspond to the usual experimental situations.  

The calculation of the viscosities (shear and bulk) and thermal conductivity can be done in molecular 
dynamics simulations using the Green-Kubo relations, which express these linear transport 
coefficients as time integrals of auto-correlation functions of microscopic 
currents \cite{hsg61,rz65,gromaz}. This formalism yields unambiguous definitions for the 
shear and bulk viscosities, applicable to both single fluids and mixtures:
\begin{subequations}
\begin{eqnarray}
&&\eta(t)=\frac{1}{6Vk_BT}\int_{0}^{t}\langle\sum_{\alpha,\beta=1;\alpha\neq\beta}^{3} 
\sigma_{\alpha\beta}(0)\sigma_{\alpha\beta}(\tau)\rangle d\tau\\
&&\zeta(t)=\frac{1}{9Vk_BT}\int_{0}^{t}\langle\sum_{\alpha,\beta=1}^{3} 
[\sigma_{\alpha\alpha}(0)-p][\sigma_{\beta\beta}(\tau)-p]\rangle d\tau
\end{eqnarray}
\label{eq:visc}
\end{subequations}
where $\hat{\mathbf{\sigma}}$ is the microscopic stress tensor:
\begin{eqnarray}
\sigma_{\alpha\beta}(\tau)=\sum_i \left[m_i v_{i\alpha}(\tau)v_{i\beta}(\tau)+
F_{i\alpha}(\tau)r_{i\beta}(\tau)\right ]
\end{eqnarray}
$(\alpha,\beta=x,y,z)$, $p$ is the pressure, and the viscosities $\eta$ and $\zeta$ 
are given by the $t\rightarrow\infty$ limits of the above relations.

The treatment of thermal transport in mixtures on the other hand is more complicated 
due to the coupling of energy and mass transport \cite{gromaz}. Since this is an important 
but many times confusing issue we discuss it briefly below for the 
present case of a binary mixture (see also the discussion in \cite{jje89}). The hydrodynamic equations 
for a binary mixture express species conservation, as well as momentum 
and entropy transport:
\begin{subequations}
\begin{eqnarray}
&&\frac{\partial\rho_a}{\partial t}+\nabla\cdot(\rho_a\mathbf{v}_a)=0\\
&&\rho\frac{\partial v}{\partial t}+\rho\mathbf{v}\cdot\nabla\mathbf{v}=-\nabla\cdot\mathbf{P}\\
&&\rho\frac{\partial s}{\partial t}+\rho\mathbf{v}\cdot\nabla s=-\nabla\cdot\mathbf{J}_s+\Theta
\end{eqnarray}
\end{subequations}
In the above equations $\rho_a$ and $\mathbf{v}_a$ $(a=1,2)$ are the (position and time dependent) mass 
densities and flow velocities of the two species, respectively; $\rho$ is the 
total mass density, $\rho=\rho_1 + \rho_2$; $\mathbf{v}$ is the center of mass (``barycentric'') velocity, 
$\mathbf{v}=\rho^{-1}(\rho_1\mathbf{v}_1+\rho_2\mathbf{v}_2)$; $\hat{\mathbf{P}}$ is the stress tensor, 
$P_{\alpha\beta}=p\delta_{\alpha\beta}-P\prime_{\alpha\beta}$, with $p$ hydrostatic pressure 
and $\hat{\mathbf{P}}\prime$ viscous stress tensor, 
$P\prime_{\alpha\beta}=[\eta(\partial v_\alpha/\partial x_\beta + \partial v_\beta/\partial x_\alpha)+
(\zeta -2\eta/3)\nabla\cdot\mathbf{v} \delta_{\alpha\beta}]$;
$s$ is the entropy density, $\mathbf{J}_s$ the entropy current 
and $\Theta$ the entropy production. The entropy current $\mathbf{J}_s$ is expressed in terms of 
heat - $\mathbf{J}_q$, and mass diffusion - $\mathbf{J}_a$, currents:
\begin{subequations}
\begin{eqnarray}
&&\mathbf{J}_s=\frac{1}{T}[\mathbf{J}_q-(\mu_1\mathbf{J}_1+\mu_2\mathbf{J}_2)]\\
&&\mathbf{J}_q=\mathbf{J}_e-(\rho e \mathbf{v} + \mathbf{P}\cdot\mathbf{v})\\
&&\mathbf{J}_a=\rho_a(\mathbf{v}_a-\mathbf{v})
\end{eqnarray}
\end{subequations}
, with $\mu_a$ chemical potential (per unit mass), $e$ total specific energy and 
$\mathbf{J}_e$ the corresponding energy current. In the framework of non-equilibrium 
thermodynamics \cite{gromaz} the heat and mass 
currents (denoted as the set $\{\mathbf{J}_\delta\}$) are 
connected to thermodynamic forces $\{\mathbf{X}_\delta\}$ 
by heat-mass linear transport coefficients 
$\{L_{\delta\gamma}\}$, $(\delta,\gamma=1,2,q)$:
\begin{equation}
\mathbf{J}_\delta=\sum_\gamma L_{\delta\gamma}\mathbf{X}_\gamma
\label{eq:ja}
\end{equation}
The entropy production $\Theta$ contains independent contributions from "vectorial" phenomena 
(heat and mass transport) - $\Theta_{\mathit{v}}$ and "tensorial" ones (momentum transport) - $\Theta_{\mathit{t}}$, 
$\Theta=\Theta_{\mathit{v}}+\Theta_{\mathit{t}}$. Both contributions assume the Onsager form, i.e. for 
heat-mass processes:
\begin{equation}
\Theta_{\mathit{v}}=\frac{1}{T}\sum_\delta \mathbf{J}_\delta\cdot \mathbf{X}_\delta
\end{equation}
while $\Theta_{\mathit{t}}=(1/T)\hat{\mathbf{P}}\prime:\nabla \mathbf{v}$. 
The Onsager reciprocity relations, $L_{\delta\gamma}= L_{\gamma\delta}$, along with 
$\mathbf{J}_1 + \mathbf{J}_2=0$ from the definition of the diffusion currents 
leave 3 independent heat-mass transport coefficients for a binary mixture, $\{L_{12}, L_{1q}, L_{qq}\}$, 
with the currents now written as
\begin{subequations}
\begin{eqnarray}
&&\mathbf{J}_1=-L_{12}(\mathbf{X}_1 - \mathbf{X}_2) + L_{1q}\mathbf{X}_q\\
&&\mathbf{J}_q=L_{1q}(\mathbf{X}_1 - \mathbf{X}_2) + L_{qq}\mathbf{X}_q
\end{eqnarray}
\end{subequations}
Three distinct sets of currents and thermodynamic forces have been discussed in detail 
\cite{gromaz,jje89}, each with different transport 
coefficients. The currents (and forces) of different sets are connected by linear 
transformations under which $\Theta_{\mathit{v}}$ is invariant and preserves its Onsager form. 
This leads to well defined and useful relations between the coefficients \cite{jje89}. 
We would like to point out that one such relation can be deduced 
without considering in detail the particular definitions of currents and forces.
We simply note that all sets use the same, physically intuitive, heat driving force, 
$\mathbf{X}_q=-\nabla T/T$, 
as well as diffusion currents $\mathbf{J}_a$ given by Eq. 9c. Since the phenomenological 
definition of the thermal 
conductivity $\lambda$ is based on the observation that in the absence of 
diffusion, i.e. $\mathbf{J}_1=0$, the 
heat current should reduce to its canonical form, $\mathbf{J}_q=-\lambda\nabla T$ 
\cite{ll00}, this yields 
\begin{eqnarray}
\lambda=\frac{1}{T}\left(L_{qq}+\frac{{L^2}_{1q}}{L_{12}}\right)
\label{eq:lambda}
\end{eqnarray}
It is worth noting that, as opposed to $L_{1q}$ and $L_{qq}$, $\lambda$ 
does not depend on the chosen set of currents and forces and moreover, it 
assumes the above form for all the sets.

For our calculations we adopt the "mainstream" choice for forces (and currents), 
\begin{subequations}
\begin{eqnarray}
&&\mathbf{X}_a=-T\nabla\left(\mu_a/T\right)\\
&&\mathbf{X}_q=-\frac{1}{T}\nabla T
\end{eqnarray}
\end{subequations}
$(a=1,2)$ \cite{jje89}, but this selection 
is not in fact arbitrary. As first discussed by Erpenbeck \cite{jje89}, 
the "mainstream" set is preferable for molecular dynamics calculations since its 
corresponding microscopic currents only depend on microscopic quantities easily 
available in simulations. The other choices 
on the other hand require the knowledge of thermodynamic quantities such as chemical 
potentials or partial enthalpies which are difficult to calculate with any accuracy 
(see also below). 

The microscopic currents for the "mainstream" set are 
\begin{eqnarray}
&&\mathbf{j}_a(\tau)=\sum_{i(a)} m_i [\mathbf{v}_i(\tau) - v_{CM}(\tau)]\\
&&\mathbf{j}_q(\tau)=\sum_i \mathbf{v}_{i}(\tau)\left\{\frac{1}{2}m_i v^2_i(\tau)+\frac{1}{2}\sum_{j\neq i}V_{ij}\left
[r_{ij}(\tau)\right]\right\}\nonumber\\
&&+\frac{1}{2}\sum_i\sum_{j\neq i}\left[\mathbf{r}_i(\tau)-\mathbf{r}_j(\tau)\right ]\mathbf{v}_i(\tau)\cdot \mathbf{F}_{ij}(\tau)
-H\mathbf{v}_{CM}(\tau)
\end{eqnarray}
where $H$ and $\mathbf{v}_{CM}$ are the enthalpy and center of mass velocity of the system, respectively. 
Since $\mathbf{v}_{CM}$ is set to zero in the simulations, $H$ does not enter in fact the 
calculations. The Green-Kubo relations for the heat-mass coefficients are:
\begin{subequations}
\begin{eqnarray}
&&L_{qq}(t)=\frac{1}{3Vk_BT}\int_{0}^{t}\langle \mathbf{j}_q(0)\cdot\mathbf{j}_q(\tau)\rangle d\tau\\
&&L_{1q}(t)=\frac{1}{3Vk_BT}\int_{0}^{t}\langle \mathbf{j}_1(0)\cdot\mathbf{j}_q(\tau)\rangle d\tau\\
&&L_{12}(t)=\frac{1}{3Vk_BT}\int_{0}^{t}\langle \mathbf{j}_1(0)\cdot\mathbf{j}_2(\tau)\rangle d\tau
\end{eqnarray}
\label{eq:heatmass}
\end{subequations}
and $\{L_{12}, L_{1q}, L_{qq}\}$ correspond to the ${t \rightarrow \infty}$ limits of the above 
relations. 

It has been sometimes remarked \cite{bh82} that the "mainstream" set does not allow a proper 
calculation of the thermal conductivity since $\lambda$ as defined by Eq. \ref{eq:lambda} may
 result from the subtraction of two large quantities leading to significant errors. 
To avoid this perceived problem these authors used a different set of currents, with microscopic 
heat current 
\begin{eqnarray}
\mathbf{j}_q^{''}=\mathbf{j}_q-(h_1\mathbf{j}_1 + h_2\mathbf{j}_2)
\label{eq:ech}
\end{eqnarray}
where $h_{1,2}$ are the partial specific enthalpies. This was expected to shift most of the 
thermal conductivity contributions to the first term of Eq. \ref{eq:lambda}, 
and therefore result in a more appropriate definition. We would like to remark that, 
if we define a time-dependent thermal conductivity
\begin{eqnarray}
\lambda(t)=\frac{1}{T}\left[L_{qq(t)}+\frac{L_{1q}(t)L_{q1}(t)}{L_{12}(t)}\right]
\end{eqnarray}
which satisfies $\lambda=lim_{t\rightarrow\infty}\lambda(t)$, it is easy to show that 
$\lambda(t)$, similarly with $\lambda$, is also invariant under such a change of currents. 
Consequently, to the extent that the thermal conductivity is defined as usual from the 
long-time "plateau" of $\lambda(t)$, using the heat current given by Eq. \ref{eq:ech} 
does not offer any real advantage. 
 
The molecular dynamics calculation of the transport coefficients relies on 
Eqs. \ref{eq:visc} and \ref{eq:heatmass}, whose integrands are easily calculated 
during simulations. 
We have performed such calculations for the systems described in 
Table 1 and also the reference (pure solvent) system. The units for viscosity 
and thermal conductivity have been chosen $(mk_BT)^{1/2}/\sigma^2$ and 
$(k_B^3T/m)^{1/2}/\sigma^2$, respectively; the time unit is $t_0=\sigma(m/k_BT)^{1/2}$.
In these units we find that the reference system has viscosities $\eta_0=1.11$, $\zeta_0=0.21$ 
and thermal conductivity $\lambda_0=4.87$. To provide an intuitive connection to the often studied 
hard sphere system we also estimate for the reference system a mean free time between 
"collisions" of $\tau\simeq 0.035$. 

Some of the autocorrelation functions (integrands) for the viscosities and thermal 
conductivity (Eqs. 6 and 17) are shown in Figs. 1 and 2
(normalized by their $t=0$ values) for the reference system and 
the colloidal system with the largest size ratio, $\sigma_c/\sigma=15$. One interesting 
feature of the shear and bulk viscosity integrands is that 
they are largely independent of the mass ratio $m_c/m$, even when it 
varies by three orders of magnitude. This feature also extends to the transport 
coefficients themselves; as shown in Fig. 6 the shear viscosity of the colloidal 
system with $m_c/m=1000$ appears to be largely similar with that of the $m_c/m=1$ 
system. Despite the fairly small colloidal "volume fraction" the time-dependent 
shear viscosity $\eta(t)$  exhibits for both 
$\sigma_c/\sigma=15$ systems a pronounced early times peak, corresponding 
to a significant viscoelastic response \cite{hm}. This effect appears much reduced 
when the diameter ratio is only slightly smaller, $\sigma_c/\sigma=10$ - Fig. 6. The 
same behavior is also observed for the bulk viscosity - Fig. 7.

The heat-mass autocorrelations - Figs. 3-5, behave qualitatively very different 
from the viscosity ones as a function of the mass of the colloidal particles. We note for 
example that they exhibit strong oscillations for light 
colloidal particles, i.e. for $m_c = m$, and a much smoother character for heavy ones, i.e. 
for $m_c/m=1000$. The thermal conductivity $\lambda(t)$ itself reflects these differences both 
at early times and as $t\rightarrow\infty$. 

In the following we would like to compare the MD results for $\eta$, $\zeta$ and $\lambda$ 
with available theories for transport in suspensions. The prediction of the transport properties 
of the present model colloidal suspension 
can proceed in principle along two different paths. The first path views the suspending liquid 
as a structureless matrix (continuum) and the colloidal 
particles as "impurities" (or "dispersed phase") with well defined properties distinct from those 
of the matrix. Then, by 
evaluating the response of the system to small, macroscopically applied fields, 
e.g. large scale temperature gradients or imposed shear flows, the equivalent, effective 
transport properties of the system can be determined (see, for example, Refs. \cite{gkb67,st02}). 
This method has a long history and is commonly known 
as effective medium theory (EMT). It has been successfully applied to both liquids and 
solids containing 
"impurities" which are large compared to any inherent matrix structure and sufficiently 
far away from each other , i.e. dilute. The predictions of these theories typically depend 
only on the volume fraction $\phi$ occupied by the dispersed phase \cite{tr95}, 
as well as "matrix" and 
"dispersed phase" properties.  For dilute enough systems the $\phi$ dependence is with 
a good approximation linear. When applied to the present case, where the colloidal particles 
considered are both "solid" 
and thermally "insulating", the transport coefficients will therefore be functions of the 
liquid "matrix" properties alone, arguably at the same pressure and temperature. 
Such relations have been in use for more than a century, and yield for the suspension 
viscosity \cite{ll00,gkb67,gkb77}, 
\begin{equation}
\eta_{eff}=\eta_0(1+\frac{5}{2}\phi)
\label{eq:einsteinvis}
\end{equation}
(Einstein's result for the viscosity of a dilute suspension), 
while for the thermal conductivity \cite{st02}
\begin{equation}
\lambda_{eff}=\lambda_0(1-\frac{3}{2}\phi)
\label{eq:maxwelltco}
\end{equation}
For the bulk viscosity the only similar result that appears 
to be available \cite{jd94} suggests that there is no contribution to the effective bulk 
viscosity to first order in $\phi$. Considering the above relations, 
the application of EMT to the present system would 
therefore seem to be straightforward and require only the value of the volume 
fraction $\phi$ occupied by the impurities, i.e. 
colloidal particles. Unfortunately this value is rather ambiguous when the modeling is done 
at the microscopic level, 
particularly when the size of the impurities is comparable with the size of the fluid 
particles, as is the case here. The problem is that EMT interprets 
$1-\phi$ as the volume fraction occupied by the fluid matrix, which is itself 
equivocal. This volume fraction may be defined as either $1-\phi_c$, or perhaps 
better for small enough $\phi_c$, as $1-\phi\prime_c$, 
where $\phi\prime_c=\phi_c(1+\sigma/\sigma_c)^3$ \cite{nmh98}; $\phi\prime_c$ accounts 
in the usual way for the actual exclusionary volume around a colloidal particle, where 
the presence of a solvent molecule is forbidden (in hard spheres language). Any 
reasonable definition of the fluid accesible volume fraction would arguably be 
bounded by these two values, $1-\phi$ and $1-\phi\prime_c$, and we therefore quote effective 
medium theory predictions corresponding to both of them. Although the difference 
between $\phi_c$ and $\phi\prime_c$ is not completely negligible, this does not 
affect our conclusions.  

The second, conceptually different treatment of transport in suspensions regards the system 
as a binary fluid mixture, i.e. it considers its microscopic, 
particle character and its detailed interparticle interactions. While no fully 
microscopic theory for the transport coefficients of either simple fluids or mixtures is available, 
the Enskog theory (ET) for the hard sphere fluid 
has proved to be successful in a significant thermodynamic domain \cite{agw}, and with suitable 
modifications has been shown to be applicable to other relevant simple fluids \cite{hmc,sb03}. 
The corresponding theory for hard 
sphere mixtures has also been rigorously derived \cite{hck83}, and tested using MD simulations 
for a number of relevant cases \cite{jje89,jje92,jje93}. Although the binary mixture studied here is modeled 
by "soft sphere"-based potentials and not hard spheres, we employ a simple scaling 
relation to estimate the relative 
values of the mixture transport coefficients with respect to those of the reference system 
at the same pressure and temperature, i.e. 
\begin{equation}
\mathbf{\Xi}(\sigma, \sigma_c, \phi, \phi_c, m, m_c)/\mathbf{\Xi_0}(\sigma, \phi_0, m)=
\mathbf{\Xi}_{ET}(\sigma, \sigma_c, \phi, \phi_c, m, m_c)/\mathbf{\Xi}_{ET0}(\sigma, \phi_0, m), 
\end{equation} 
where $\mathbf{\Xi}$ stands for $\eta$, $\zeta$ or $\lambda$, and $0$ denotes the reference (pure solvent) system. 
The Enskog theory hard sphere 
results (denoted above by $ET_0$) are well known \cite{agw}, while the Enskog mixture theory relations 
can be found in \cite{hck83}; 
they are too complicated to be meaningfully quoted here. We 
used the second Enskog approximation \cite{hck83} and tested our numerical implementation 
against the values quoted in \cite{jje93}.

We now proceed to compare the MD simulation results for shear and bulk viscosities and thermal 
conductivity with the predictions of these two theories - see Table 2. (We note first 
that the quoted MD values carry statistical error bars of approximately 10-15\%, which 
should also encompass 
small deviations due to the neglect of long time tails \cite{jje93,jje92}.) 
The MD calculated 
shear viscosity is fairly well reproduced by the effective medium theory Eq. \ref{eq:einsteinvis} 
for both large and small colloidal masses. The Enskog predictions on the other hand appear 
to deviate significantly from the MD values, particularly for the larger colloidal particles. 
Surprisingly, the opposite seems to hold for the bulk viscosity, which is found to be 
much larger 
than that of the reference system, in good agreement with the Enskog results but not EMT. 
The apparent 
failure of EMT for the bulk viscosity may signal that the theory needs significant 
corrections for "soft" colloidal particles, or perhaps even for standard hard cores. 
The comparison of the MD values and theoretical predictions 
for the thermal conductivity yields the more interesting results. 
For heavy colloidal particles both the EMT and Enskog predictions appear to be in reasonable 
agreement with the MD simulations. On the other hand, if the colloidal particles are light, 
$m_c/m=1$, particularly if the size ratio is also large, the thermal conductivity is found to 
be significantly bigger than the EMT prediction. In fact, for $\sigma_c/\sigma=15$ 
it appears that by adding to the solvent (at constant pressure) thermally insulating but 
rather small and light colloidal particles, the thermal conductivity is significantly 
($\simeq 50\%$) enhanced over that of the reference, pure solvent! 
The Enskog theory predictions also appear to roughly reproduce this trend. 

In conclusion, we performed MD calculations of the viscosities and thermal conductivity 
of a model colloidal suspension with colloidal particles approximately only one order 
of magnitude larger than the solvent molecules (``nanocolloidal'' suspension). We compare 
the calculated values with those corresponding to the pure host liquid (``solvent'') at the 
same pressure and temperature, and test their ratio against two conceptually distinct theories: 
effective medium theory (EMT) and Enskog transport theory. 
The results suggest that, quite remarkably, the standard EMT 
remains well applicable for predicting both the shear viscosity and thermal conductivity of such 
suspensions when the colloidal particles have a "typical" mass, 
i.e. $m_c/m\sim (\sigma_c/\sigma)^3$. Somewhat puzzling, the available EMT result for the suspension bulk 
viscosity (essentially identical here with the reference liquid value) 
fails to reproduce the MD calculations; this may indicate that some revised theory 
is necessary. Estimates of the transport coefficients based on the Enskog transport theory 
are less conclusive, but appear to suggest that when applied to systems as 
the ones studied 
here the theory is rather inaccurate for the shear viscosity, although it may remain satisfactory 
for the bulk viscosity and the thermal conductivity.
For extremely light colloidal particles, i.e. $m_c/m\sim 1$, we find a significant 
thermal conductivity enhancement over the EMT predictions, which is roughly 
reproduced by the Enskog mixture theory. This observed mass dependence appears to distinguish 
in principle, from a heat conduction point of view, liquid suspensions from their solid 
phase counterparts to which EMT is most often applied. The effect may be 
perhaps attributed to the solvent ``stirring'' action of the Brownian colloidal particles' motion, 
posited for example in \cite{pbp05}. However interesting, this behavior does not appear to be of 
much consequence for common, realistic suspensions, including nanocolloidal ones or 
so-called nanofluids, since for such systems the colloidal particles are dense objects with 
a mass much larger than that of the solvent molecules, and are therefore ``typical'' in the sense 
noted above. On the other hand the effect may be relevant for ``bubbly'' 
liquids, where the density of the host liquid is much larger than that of the suspended 
``particles'' (gas bubbles). The present predictions would pertain in principle to nanobubbles-rich 
fluids, although we are not aware of any available experimental results that can 
either confirm or refute their applicability.
 
I thank one of the referees for pointing out the possible relevance of the simulations 
with light colloidal particles to ``bubbly'' liquids. 
This work was performed under the auspices of the U. S. Department of Energy by 
University of California Lawrence Livermore National Laboratory under Contract 
No. W-7405-Eng-48.

\newpage
\begin{table}
\caption{\label{tab:systems}Colloidal systems studied.}
\begin{ruledtabular}
\begin{tabular}{ccccccc}
&$\sigma_c/\sigma$ &$m_c/m$ &$N_c$  &$\phi_c$ &$N$ &$\phi$ \\
\hline
&10 &1 &25 &0.0751 &127675 &0.384 \\
&10 &1000 &25 &0.0751 &127675 &0.384 \\
&15 &1 &20 &0.0824 &312638 &0.382 \\
&15 &1000 &20 &0.0824 &312638 &0.382 \\
\end{tabular}
\end{ruledtabular}
\end{table}

\newpage
\begin{table}
\caption{\label{tab:results}Comparison of the shear and bulk viscosities, and thermal 
conductivity (top to bottom) of the colloidal suspensions from MD simulations, 
effective medium theory (EMT) and Enskog theory for mixtures. The two EMT numbers 
correspond to volume fractions $\phi_c$ and $\phi\prime_c$ (see text).}
\begin{ruledtabular}
\begin{tabular}{ccccccc}
&$\sigma_c/\sigma$ &$m_c/m$ &$MD $ &$EMT$ &$Enskog$ \\
\hline
&10 &1 &1.12 &1.19-1.25 &1.34 \\
&10 &1000 &1.16 &1.19-1.25 &1.59 \\
&15 &1 &1.15 &1.21-1.25 &1.54 \\
&15 &1000 &1.21 &1.21-1.25 &1.86 \\
\hline
&10 &1 &1.78 &1 &1.44 \\
&10 &1000 &1.95 &1 &1.83 \\
&15 &1 &1.86 &1 &1.71 \\
&15 &1000 &2.20 &1 &2.22 \\
\hline
&10 &1 &0.95 &0.85-0.89 &1.26 \\
&10 &1000 &0.85 &0.85-0.89 &0.91 \\
&15 &1 &1.52 &0.85-0.88 &1.40 \\
&15 &1000 &0.80 &0.85-0.88 &0.91 \\
\end{tabular}
\end{ruledtabular}
\end{table}

\newpage
\begin{figure}
\includegraphics{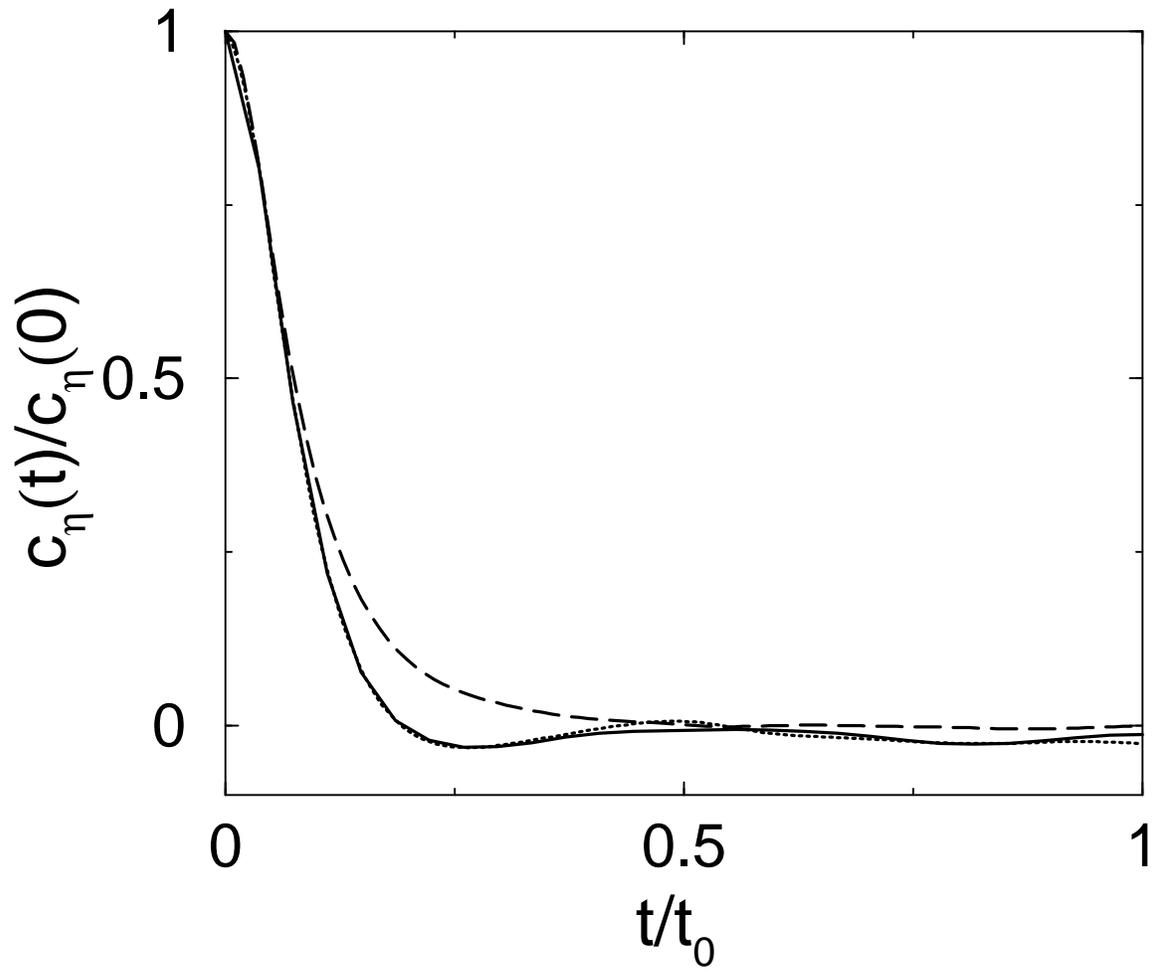}
\caption{Shear viscosity autocorrelation function (normalized by its $t=0$ value) 
for the reference system (dashed line), and mixtures with $\sigma_c/\sigma = 15$; 
$m_c/m = 1000$ (solid line) and $m_c/m = 1$ (dotted line).}
\label{fig:fig1}
\end{figure}

\newpage
\begin{figure}
\includegraphics{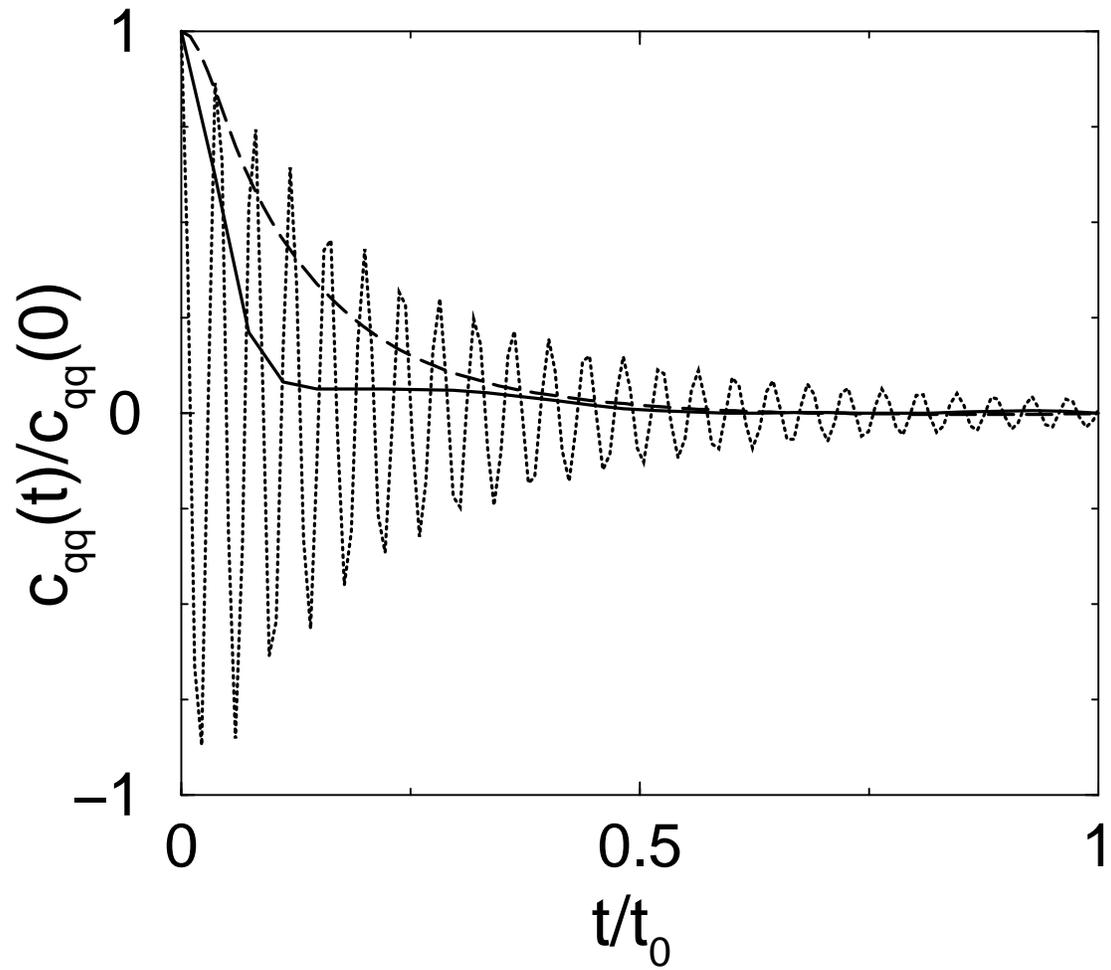}
\caption{Autocorrelation function (normalized by its $t=0$ value) for the $L_{qq}$ 
coefficient of the reference system (dashed line), and mixtures with 
$\sigma_c/\sigma = 15$; $m_c/m = 1000$ (solid line) and $m_c/m = 1$ (dotted line).}
\label{fig:fig5}
\end{figure}

\newpage
\begin{figure}
\includegraphics{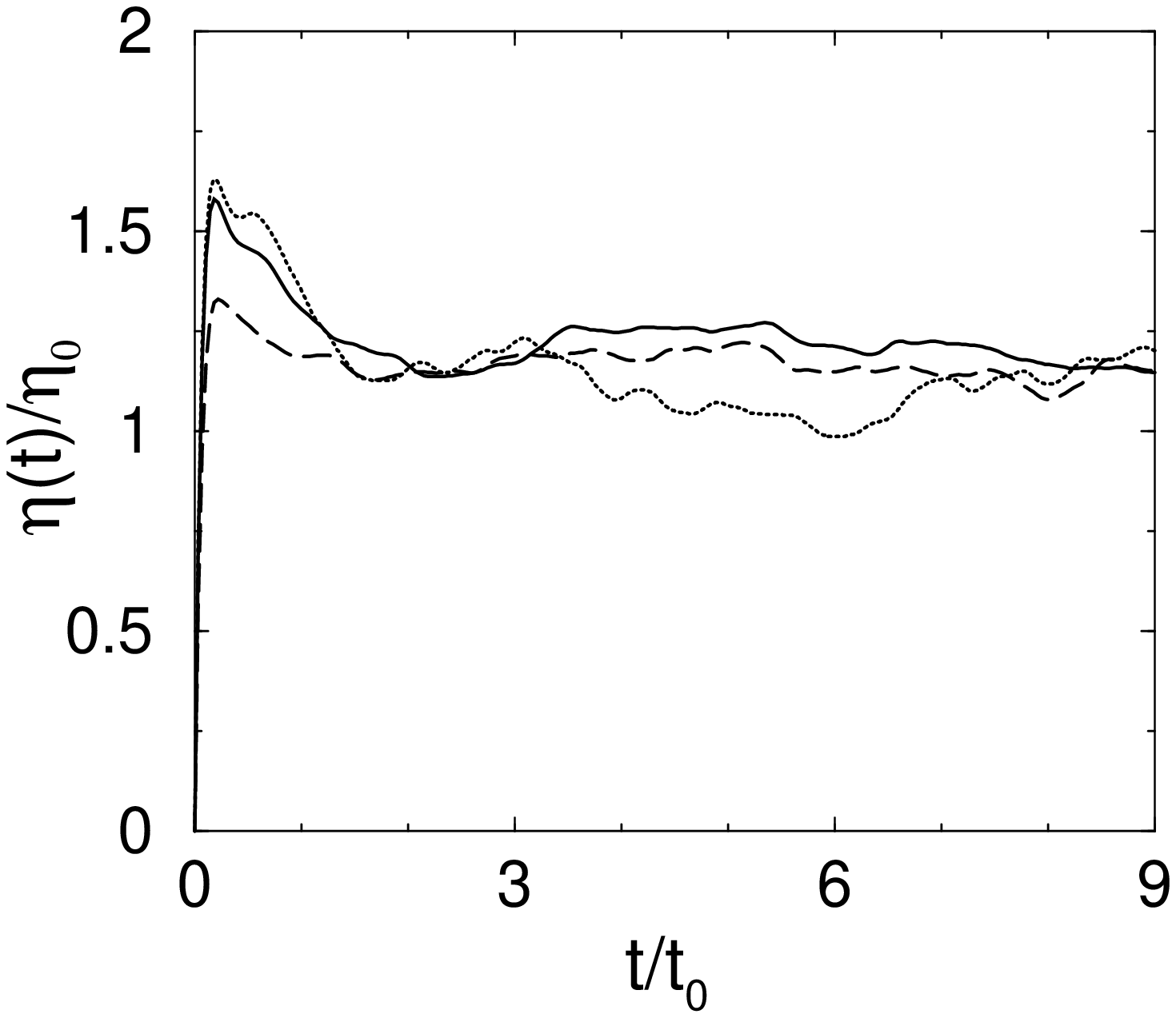}
\caption{Time dependent shear viscosity (normalized by the reference system value) for mixtures 
with $\sigma_c/\sigma = 15$, $m_c/m = 1000$ (solid line), 
$\sigma_c/\sigma = 10$, $m_c/m = 1000$ (dashed line), and 
$\sigma_c/\sigma = 15$, $m_c/m = 1$ (dotted line).}
\label{fig:fig6}
\end{figure}

\newpage
\begin{figure}
\includegraphics{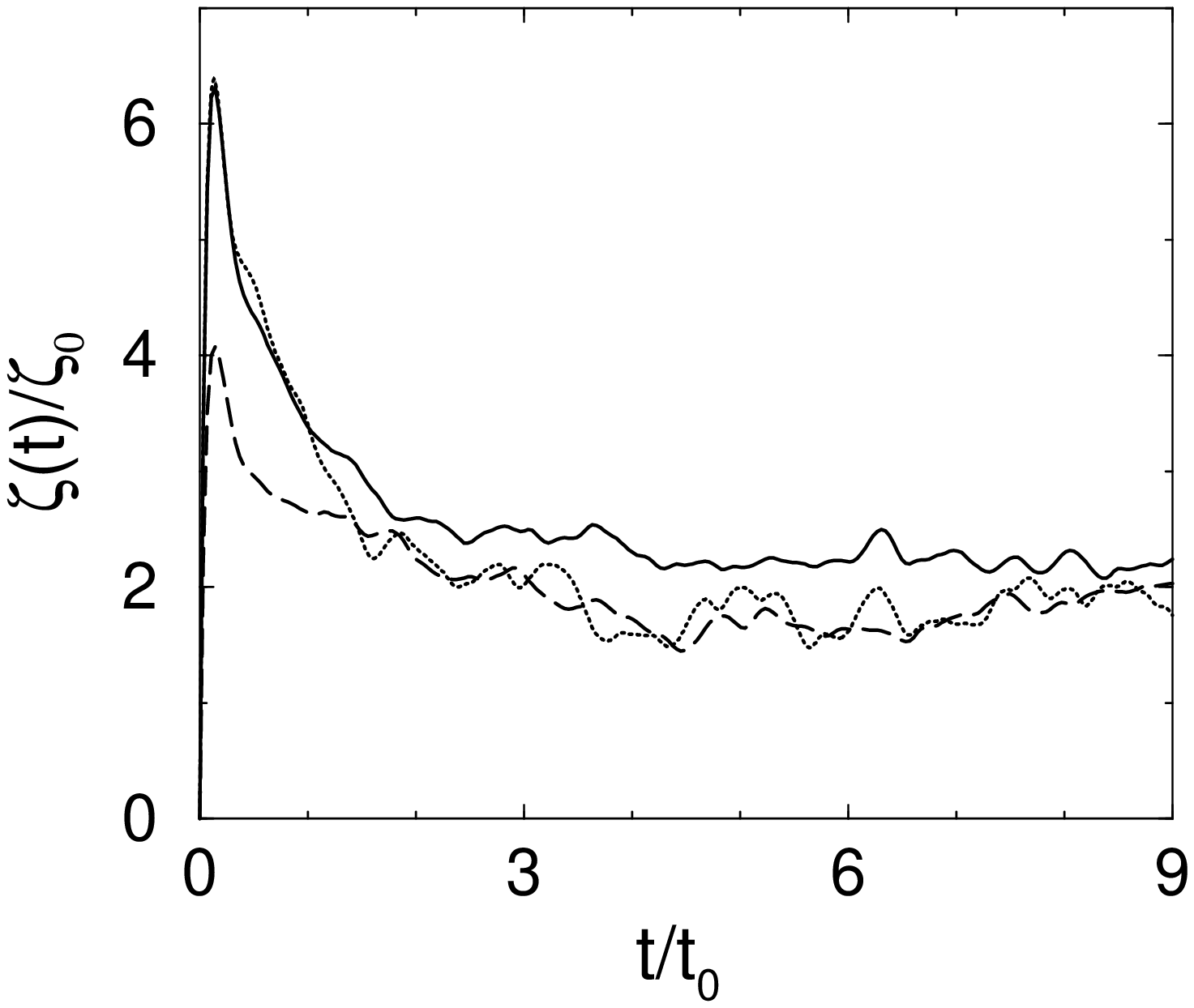}
\caption{Time dependent bulk viscosity (normalized by the reference system value) for mixtures 
with $\sigma_c/\sigma = 15$, $m_c/m = 1000$ (solid line), 
$\sigma_c/\sigma = 10$, $m_c/m = 1000$ (dashed line), and 
$\sigma_c/\sigma = 15$, $m_c/m = 1$ (dotted line).}
\label{fig:fig7}
\end{figure}

\newpage
\begin{figure}
\includegraphics{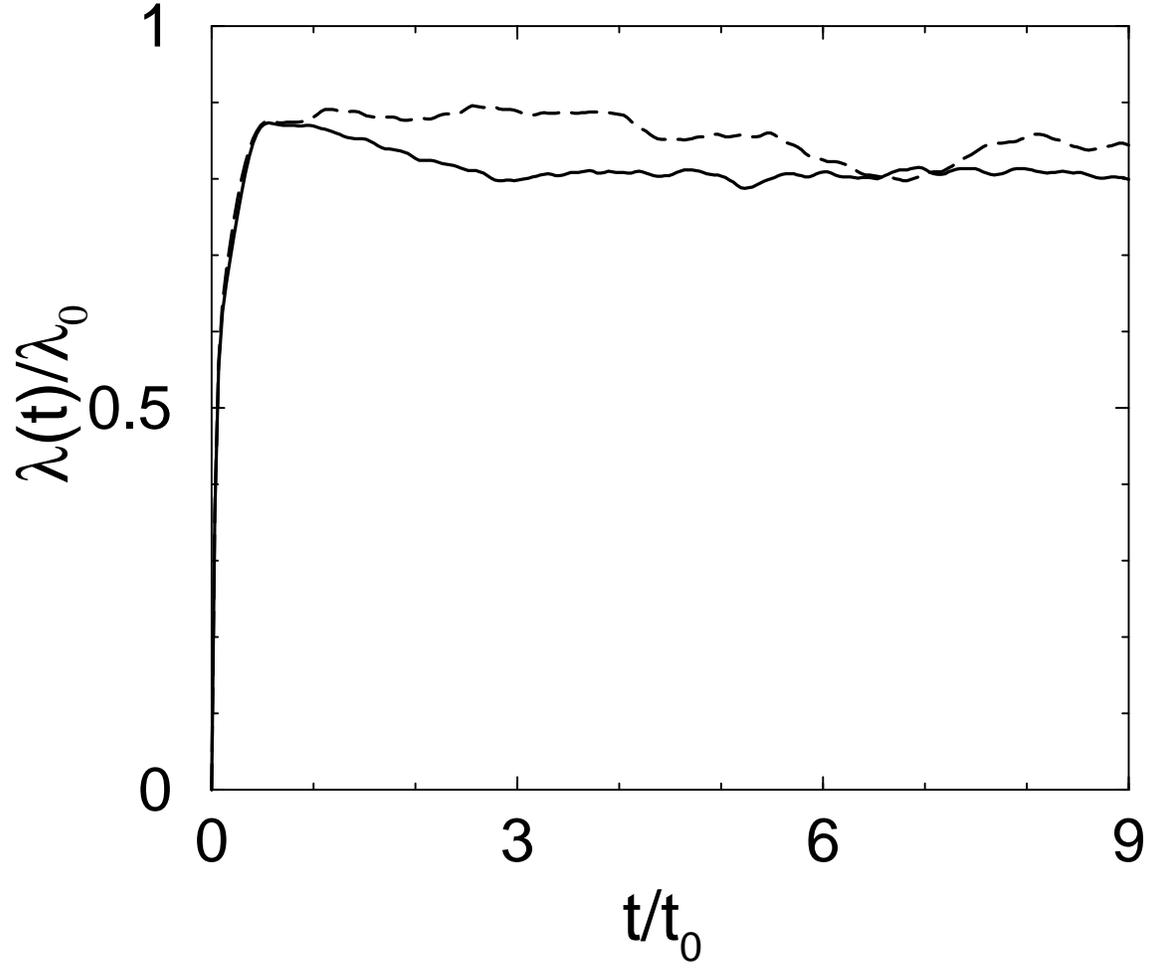}
\caption{Time dependent thermal conductivity (normalized by the reference system value) for mixtures 
with $\sigma_c/\sigma = 15$, $m_c/m = 1000$ (solid line), 
and $\sigma_c/\sigma = 10$, $m_c/m = 1000$ (dashed line).}
\label{fig:fig8}
\end{figure}

\newpage
\begin{figure}
\includegraphics{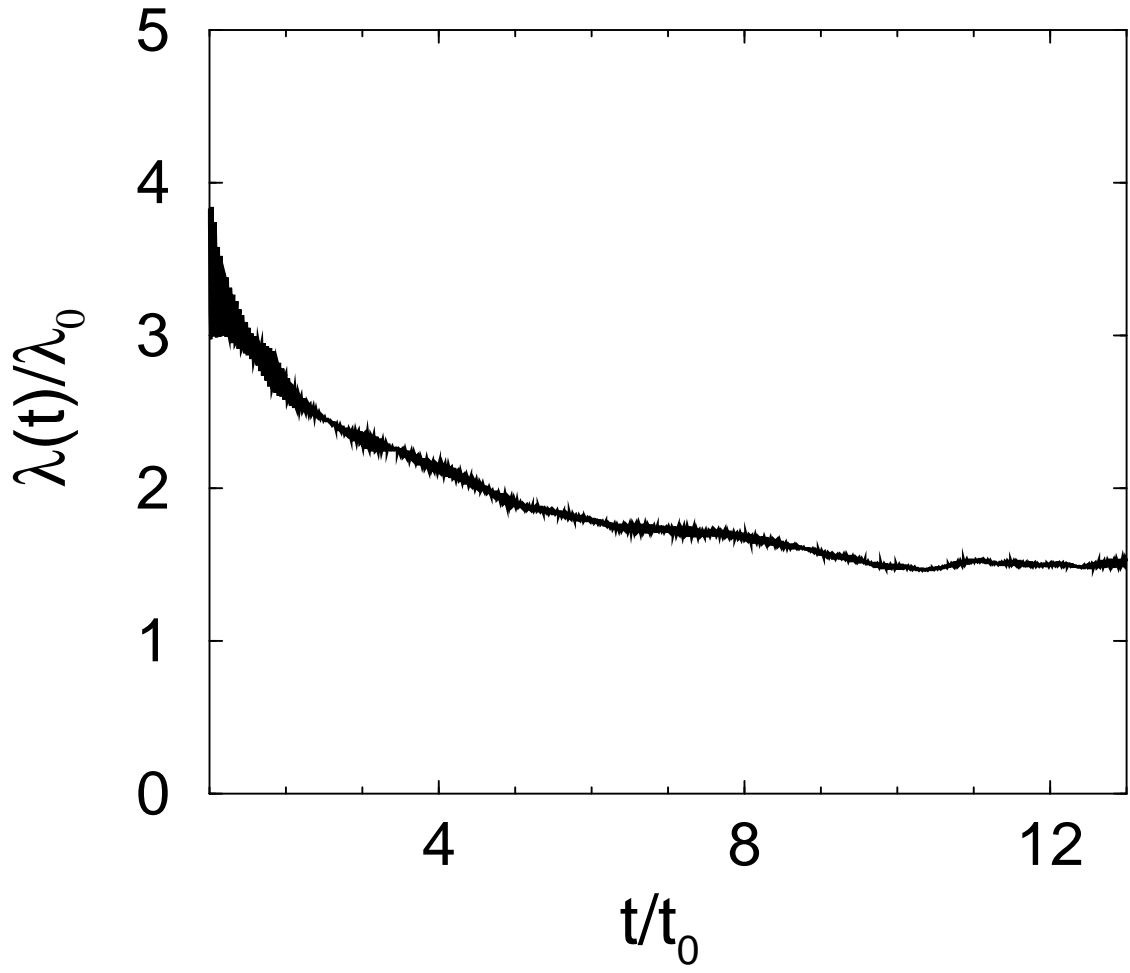}
\caption{Time dependent thermal conductivity (normalized by the reference fluid value) 
for the mixture with $\sigma_c/\sigma = 15$ and $m_c/m = 1$.}, 
\label{fig:fig9}
\end{figure}

\end{document}